\documentstyle[preprint,tighten,aps,psfig]{revtex}
\begin{document}
\preprint{TTP96-05, hep-ph/9603261}
\draft
\title{Two-loop QCD corrections to $b\to c$ transitions at zero recoil}
\author{Andrzej Czarnecki}
\address{Institut f\"ur Theoretische Teilchenphysik, 
Universit\"at Karlsruhe,\\ D-76128 Karlsruhe, Germany}
\maketitle

\begin{abstract}
Complete two-loop QCD corrections to $b\to c$ transitions are
presented in the limit of zero recoil.  Vector and axial-vector
coefficients $\eta_{A,V}$ are calculated analytically in the limit of
equal beauty and charm masses, and a series appoximation is obtained
for the general mass case.  $\eta_A$ is crucial for the determination
of the absolute value of the Cabibbo-Kobayashi-Maskawa matrix element
$V_{cb}$.  The two-loop effects enhance the one-loop corrections by
21\%, removing a major theoretical uncertainty in the value of
$|V_{cb}|$.  Including two-loop QCD effects and 
previously neglected electroweak and Coulomb
corrections we find $|V_{cb}| = 0.0379 \pm 0.0020({\rm stat})
\pm 0.0025({\rm syst}) \pm 0.0011({\rm theory})$.
\end{abstract}

\vspace*{10mm}

Elements of the Cabibbo-Kobayashi-Maskawa (CKM) matrix are fundamental
input parameters of the Standard Model.  Their precise measurements
have been the subject of vast experimental efforts and will remain
prominent issues in forthcoming projects, most notably the
B-factories.  The values of CKM matrix elements determine the sides of
the unitarity triangle and their precise knowledge is essential for
the understanding of the origin of CP violation, a major puzzle of the
Standard Model.

One of the directly measurable CKM parameters is the absolute value of
$V_{cb}$.  The experimental value can be extracted from decays of $B$
mesons produced on the $\Upsilon(4)$ resonance ({\sc ARGUS}
Collab.~\cite{ARGUS93}, {\sc CLEO} Collab.~\cite{CLEO95}) or on the
$Z$ resonance ({\sc ALEPH} Collab.~\cite{ALEPH95}, {\sc DELPHI}
Collab.~\cite{DELPHI96}).

$|V_{cb}|$ can be obtained either from the total width of semileptonic
$B$ decays or from the zero-recoil extrapolation of the exclusive
decay spectrum of $B\to D^* l\bar\nu$, where $l$ is an electron or
muon (see \cite{Mannel95} for a recent review).  The merits of both
methods and theoretical uncertainties have been discussed in
ref.~\cite{Shifman94}.  The inclusive approach has the advantage of
larger experimental statistics; the inherent theoretical error is
mainly due to inaccurate knowledge of the quark masses which enter the
decay width formula.  This theoretical uncertainty already dominates
the experimental error and it is not obvious that it can be
significantly improved (see, however, a discussion in
ref.~\cite{Bigi95} and references therein).

The exclusive method, on the
other hand, benefits from recent advances \cite{Shifman94,Neubert95}
in the heavy quark effective theory (HQET)
\cite{Isgur89,Isgur90,Eichten90,SV88}.  It has been used to obtain the
latest experimental result
\begin{eqnarray}
|V_{cb}| = 0.0385 \pm 0.0021({\rm stat})
 \pm 0.0025({\rm syst})
\pm 0.0017({\rm theory}) \quad \mbox{(ref.~\cite{DELPHI96})}.
\label{eq:exp}
\end{eqnarray}
The exclusive method can be summarized as follows:
The recoil spectrum of the $B$ meson decay is
\begin{eqnarray}
{{\rm d} \Gamma(B\to D^* l\bar \nu) \over {\rm d}w}
&=& f(m_B,m_{D^*}) |V_{cb}|^2 {\cal F}^2 (w)
\nonumber\\
&&\qquad\times\left( 1+ {\alpha\over \pi}\ln {M_Z\over m_B} \right) 
,
\label{eq:spectrum}
\end{eqnarray}
where $w$ is the product of the four-velocities of the $B$ and $D^*$
mesons, and $f$ is a known (see e.g.~\cite{Neubert95}) function of
masses of observable particles (rather than of quark masses).  HQET
offers a model-independent value of the hadronic matrix element for
the decay $B\to D^* l\bar \nu$ at zero recoil, ${\cal F}$(1), up to
perturbative corrections, to be subsequently discussed.  This point is
not directly accessible in the experiment due to the vanishing phase
space.  Fortunately, $|V_{cb}|^2{\cal F}^2 (1)$ can be deduced by
extrapolating the measured values at non-zero recoil, and, given the
theoretical prediction for ${\cal F} (1)$, the value of $|V_{cb}|$ can
be obtained.  The last factor in eq.~(\ref{eq:spectrum})
approximates the electroweak corrections \cite{Sirlin82}.  In case of
neutral $B$ decays an additional factor $(1+\pi\alpha)$ should be
included in eq.~(\ref{eq:spectrum}) (see \cite{Atwood90} and
references therein.)  It represents a 2.3\% enhancement of the rate
due to the final state interaction between the lepton and the charged
$D^*$ meson.

 In contrast to the decay into a pseudoscalar $D$ meson,
the prediction of HQET for $B\to D^*$ transition is free from
$1/m_{b,c}$ corrections \cite{Neubert91} by virtue of Luke's theorem
\cite{Luke90}.      The form factor ${\cal F} (1)$ can be written as
\begin{eqnarray}
{\cal F}(1) =  \eta_A \left(1+\delta_{1/m^2}\right),
\end{eqnarray}
where $\eta_A=1+{\cal O}(\alpha_s)$ describes the perturbative QCD
corrections. 
The mass corrections $\delta_{1/m^2}$ of 
order $1/m_Q^2$ have been examined \cite{Shifman94,Neubert94}.  They
are estimated \cite{Neubert94} 
to decrease the form factor by $(5.5\pm 2.5)\%$
and their error is responsible for approximately half of the
theoretical uncertainty in the value of $|V_{cb}|$ quoted in
eq.~(\ref{eq:exp}).

The large remaining theoretical uncertainty is due to the unknown 
perturbative QCD corrections.  The one-loop corrections are formally
identical to those calculated in the context of muon decay
\cite{beh56}.  For the heavy quark
decays they give \cite{Paschalis83,Close84,SV88} 
\begin{eqnarray}
\eta_A &=& 1 
- {\alpha_s\over \pi} C_F  
\left( {3\over 4}{2-\delta \over \delta}\ln(1-\delta)  
+ 2\right) 
\nonumber \\&&
+ \left({\alpha_s\over \pi}\right)^2 C_F \eta^{(2)}_A 
+{\cal O}(\alpha_s^3),
\end{eqnarray}
where the dimensionless variable $\delta=1-m_c/m_b$ denotes the
relative difference of quark masses, and $\eta^{(2)}_A$ represents the
two-loop QCD corrections.  The latter have been subject of vigorous
controversy over the last few years.  In the absence of an exact
calculation, a renormalization group analysis has been performed
\cite{Neubert92} but its validity has been questioned in view of the
small size of the logarithm of the mass ratio $m_b/m_c$
\cite{Uraltsev95,Shifman95}.  The need for a full two-loop calculation
of $\eta^{(2)}_A$ has been emphasized by many authors
\cite{Neubert92,Uraltsev95,Shifman95,Buras95}.  The purpose of this
paper is to provide this correction.

A calculation of QCD (or even QED) two-loop corrections to a fermion
decay is in general very difficult.  A full calculation has never been
done, neither for the muon nor for a quark.  However, it is at present
possible to perform such an analysis at least at the zero recoil
point.  The advantage of this particular kinematical point is
twofold. First, because of the phase-space suppression, there is no
room for real radiation.  Therefore, the virtual corrections are
infrared finite.  Second, the four vectors of the decaying and final
quarks are parallel.

These two features of the zero recoil configuration allow an exact
analytic solution in the case of equal masses $m_b$ and $m_c$; in the
general mass case one can construct an approximate solution in the
form of a power series in the relative mass difference $\delta$. In
addition, the solution has a very useful symmetry with respect to the
exchange $m_b\leftrightarrow m_c$.  This non-trivial symmetry, valid
only at zero recoil, helps to extract maximum information from the
approximating series by accelerating its convergence.

The two-loop QCD diagrams relevant to this calculation are shown in
fig.~1.  The Lorentz structure of the decay vertex is
\begin{eqnarray}
\Gamma_\mu = \gamma_\mu (1-\gamma_5),
\end{eqnarray}
and the vector and axial-vector parts of this coupling are modified in
different ways by the QCD corrections;  at zero recoil they are 
parametrized by two functions, $\eta_{V,A}$
\begin{eqnarray}
\gamma_\mu \to  \eta_V \gamma_\mu,
\qquad
\gamma_\mu\gamma_5 \to  \eta_A \gamma_\mu\gamma_5.
\end{eqnarray}
Although for the decay  $B\to D^*l\bar\nu$  only the axial part is
relevant, for the sake of completeness we also compute $\eta_V$.
The one-loop result for $\eta_V$ is known
\cite{beh56,Paschalis83,Close84,SV88}.  
\begin{eqnarray}
\eta_V &=& 1 
- {\alpha_s\over \pi} C_F  \left( {3\over 4}{2-\delta\over \delta}
\ln(1-\delta)  +
{3\over 2}\right) 
\nonumber\\ &&
+ \left({\alpha_s\over \pi}\right)^2 C_F \eta^{(2)}_V 
+{\cal O}(\alpha_s^3).
\end{eqnarray}
The one-loop QCD corrections contain only one color
structure, proportional to $C_F$.  At the two-loop  level 
it is convenient to divide up the 
functions $\eta_{A,V}^{(2)}$ into four parts proportional to
various SU(3) factors (an overall factor $C_F$ has been factored out):
\begin{eqnarray}
\eta_{A,V}^{(2)} &=& C_F \eta_{A,V}^F + (C_A -2 C_F)\eta_{A,V}^{AF}
\nonumber\\
&&
+T_R N_L \eta_{A,V}^L +T_R  \eta_{A,V}^H.
\label{eq:colors}
\end{eqnarray}
For a general SU(N) group $C_A=N$; $C_F=(N^2-1)/2N$; $T_R=1/2$.  $N_L$
denotes the number of the light quark flavors whose masses can be
neglected.  The last term contains contributions of the massive quark
loops, with $b$ and $c$ quarks.  We neglect the top quark; its impact
is suppressed by a factor $\sim m_b^2/m_t^2$.  

Among the eight coefficient functions in eq.~(\ref{eq:colors})
$\eta_{A,V}^L$ are already known \cite{Neubert95beta}.  They
corresponds to our diagram (f) in fig.~1 with a massless fermion in
the loop.  In the $\overline{\rm MS}$ scheme (with
$\mu=\sqrt{m_bm_c}$), adopted also in the present work, they read
\begin{eqnarray}
\eta_A^L &=& {5\over 24}\left[ {2-\delta\over \delta} \ln(1-\delta)+
{44\over 15}\right],
\nonumber\\
\eta_V^L &=& {1\over 24}\left[ {2-\delta\over \delta} \ln(1-\delta)+2
\right].
\end{eqnarray}
The remaining six functions can be calculated exactly in the case of
equal masses $m_b$ and $m_c$.  In this limit the momenta of the
leptons in the final state vanish and the vertex function becomes a
two-point function with a zero momentum insertion.  Such
propagator-like on-shell functions are  known; a systematic
method of their evaluation has been worked out in
ref.~\cite{gbgs90,GrayPhD}; 
the underlying idea is the integration by parts
method \cite{che81}.
This method has greatly simplified
the two-loop QED calculation of $g-2$ \cite{shell2,czakam92}.
In ref.~\cite{shell2} the computation of two-loop functions with a low
number of zero momentum insertions has been automated.  For the
purpose of the present calculation  a new implementation of the recurrence
algorithm \cite{bro91a} was necessary; this is because of the
necessity of computing two-loop functions with large number of zero
momentum insertions.

In order to go beyond the $m_b=m_c$ limit we use the variable $\delta$
as an expansion parameter. In the real world $m_b$ and $m_c$ are far
from being equal; for the purpose of this work we take $m_b = 4.8$ GeV
and $m_c =1.44$ GeV which yields $\delta= 0.7$.  Coefficients of the
expansion of $\eta_{A,V}$ in $\delta$ are two-point on-shell functions
which can be computed using recurrence relations.  In order to ensure
good numerical accuracy we have computed ten terms in the $\delta$
expansion for all diagrams in fig.~1.  The analytic computation of the
resulting integrals was feasible only thanks to the latest
achievements in symbolic manipulation programs \cite{form}.

The results we obtained are symmetric with respect to the exchange
$m_b\leftrightarrow m_c$, or $\delta \to -\delta/(1-\delta)$.  The
resulting fact that every other term in the $\delta$ expansion can be
obtained from the earlier terms provides a strong consistency check of
our procedures.  On the other hand, it is possible to rewrite the
series expansion in a manifestly symmetric form.  For this purpose we
introduce a variable invariant with respect to $m_b\leftrightarrow
m_c$, $\rho \equiv {\delta^2/ ( 1-\delta )}$.
In terms of this variable our results can be written as a rapidly
convergent series and we can safely neglect terms ${\cal O}(\rho^5)$.
Quadratic and linear logarithmic terms, implicitly present in this
expansion \cite{Falk90,Ji91,BroGro91,Neubert92}, 
do not spoil the convergence because $\ln (m_b/m_c)$ is of
order unity.

For the axial-vector function $\eta_A^{(2)}$ we find:
\begin{eqnarray}
\lefteqn{\eta^{AF}_{A} =
 - {143\over 144} - {1\over 12}\pi^2 
 + {1\over 6}\pi^2\ln 2  - {1\over 4}\zeta(3)}
\nonumber\\
&& \;\;\;
 + \rho \left( {29\over 576} + {55\over 1728}\pi^2 
 - {1\over 16}\pi^2\ln 2 + {3\over 32}\zeta(3)\right) 
\nonumber\\
&& \;\;\;
 + \rho^2 \left(  - {2509\over 17280} + {121\over 8640}\pi^2\right)
 + \rho^3 \left( {43\over 22680} - {67\over 967680}\pi^2\right)
\nonumber\\
&& \;\;\;
 + \rho^4 \left(  
      - {17933\over 50803200} + {143\over 11612160}\pi^2\right)
\nonumber\\
\lefteqn{\eta^{F}_A =
 - {373\over 144} + {1\over 6}\pi^2}
\nonumber\\
&& \;\;\;
 + \rho \left( {377\over 576} - {1\over 24}\pi^2\right)
 + \rho^2 \left(  - {29\over 648} + {1\over 360}\pi^2\right)
\nonumber\\
&& \;\;\;
 + \rho^3 \left( {227\over 37800} - {1\over 2520}\pi^2\right)
 + \rho^4 \left(  - {649\over 672000} + {1\over 15120}\pi^2\right)
\nonumber\\
\lefteqn{\eta^{H}_A =
  {115\over 18} - {2\over 3}\pi^2}
\nonumber\\
&& \;\;\;
 + \rho \left( {529\over 72} - {107\over 144}\pi^2\right)
 + \rho^2 \left(  - {337\over 144} + {137\over 576}\pi^2\right)
\nonumber\\
&& \;\;\;
 + \rho^3 \left(  - {255313\over 75600} + {197\over 576}\pi^2\right)
 + \rho^4 \left(  - {1957573\over 3175200} + {1\over 16}\pi^2\right)
\nonumber\\
\label{eq:mainA}
\end{eqnarray}
For the corrections to the vector current we find:
\begin{eqnarray}
\lefteqn{\eta^{AF}_V =
  \rho \left( {377\over 576}  - {3\over 64}\pi^2 
 - {1\over 16}\pi^2\ln 2 + {3\over 32}\zeta(3)\right)} 
\nonumber\\
&& \;\;\;
 + \rho^2 \left(  - {107\over 5760} + {1\over 288}\pi^2\right)
 + \rho^3 \left( {41\over 10080} - {31\over 46080}\pi^2\right)
\nonumber\\
&& \;\;\;
 + \rho^4 \left(  
     - {13927\over 16934400} + {169\over 1290240}\pi^2\right)
\nonumber\\
\lefteqn{\eta^{F}_V =
  \rho \left( {553\over 576} - {5\over 72}\pi^2\right)
 + \rho^2 \left(  - {227\over 4320} + {1\over 360}\pi^2\right)}
\nonumber\\
&& \;\,
 + \rho^3 \left( {251\over 33600} - {1\over 2520}\pi^2\right)
 + \rho^4 \left(  - {7537\over 6048000} + {1\over 15120}\pi^2\right)
\nonumber\\
\lefteqn{\eta^{H}_V =
  \rho \left( {197\over 72} - {13\over 48}\pi^2\right)
 + \rho^2 \left(  - {701\over 720} + {19\over 192}\pi^2\right)}
\nonumber\\
&& \;\;\;
 + \rho^3 \left(  - {2851\over 1008} + {55\over 192}\pi^2\right)
 + \rho^4 \left(  - {93227\over 151200} + {1\over 16}\pi^2\right)
\nonumber\\
\label{eq:mainV}
\end{eqnarray}

We note that all QCD contributions to $\eta_V$ vanish at $m_b=m_c$
($\rho=0$), in consequence of vector current conservation.

So far we have not discussed the renormalization procedure which led
to the results in eq.~(\ref{eq:mainA},\ref{eq:mainV}).  For the
external quark legs we used the two-loop quark wave function
renormalization constant computed in \cite{bgs91}.
Vanishing of the terms independent of $\rho$ in eq.~(\ref{eq:mainV})
serves as an independent check of the complicated expressions given in
\cite{bgs91}.  The diagrams (b1) and (b3) require  mass
counterterms.  For these we adopted the on-shell condition.  Our
results are therefore in terms of the pole masses $m_b$ and $m_c$.
For the coupling constant renormalization we used the $\overline{\rm
MS}$ condition with the renormalization scale at the geometric mean
mass $\mu=\sqrt{m_bm_c}$.  It must be noted that the symmetry
$m_b\leftrightarrow m_c$ is in general valid only for the
unrenormalized diagrams.  If the coupling constant were normalized at
a different $\mu$ the final result~(\ref{eq:mainA},\ref{eq:mainV})
would not be symmetric.  The geometric mean point is the only one
preserving that symmetry.

Numerically, the two-loop corrections  evaluate to
\begin{eqnarray}
\lefteqn{\eta_A^{(2)} = -0.586(2)\times C_F - 0.909(2) \times (C_A-2C_F)}
\nonumber \\ &&\qquad
+0.145 \times T_RN_L -0.155(4)\times T_R = -0.944(5),
\nonumber \\
\lefteqn{\eta_V^{(2)} = 0.395(2)\times C_F - 0.168(2) \times (C_A-2C_F)}
\nonumber \\ &&\qquad
-0.010 \times T_RN_L +0.107(2)\times T_R = 0.509(5).
\end{eqnarray}
It is interesting to compare these results with an estimate based on
the subset of corrections of order $\alpha_s^2\beta_0$ with
$\beta_0=11-{2\over 3}n_f$.  With $n_f=4$ one
gets \cite{Neubert95beta}
\begin{eqnarray}
\eta_A^{(2)} = -0.908,\quad \eta_V^{(2)} = 0.061.
\label{eq:beta}
\end{eqnarray}
In the axial-vector case the agreement with the full two-loop
calculation is quite good.  The estimate fails badly in the vector
case; this is probably because of an accidental numerical cancellation
in (\ref{eq:beta}) which makes the estimate of $\eta_V^{(2)}$ very small.
The full one and two-loop corrections tend to give corrections to
$\eta_V$ with  approximately half the magnitude of those to $\eta_A$.

Adopting $\alpha_s(\sqrt{m_bm_c})=0.24$,
our full two loop calculation leads to the total values of $\eta_{A,V}$
\begin{eqnarray}
\eta_A &=& 1- 0.033-0.007 = 0.960\pm 0.007,
\nonumber \\
\eta_V &=& 1 + 0.018 + 0.004 = 1.022\pm 0.004.
\end{eqnarray}
Since the perturbative series in QCD is asymptotic, the uncertainty in
the values of $\eta_{A,V}$ has been estimated by the size of the last
computed terms.  The central value we obtain for $\eta_A$ is
consistent with the value given by Neubert,
$\eta_A({\mbox{ref.~\cite{Neubert95}}})=0.965\pm 0.020$, which was
adopted in the recent experimental studies.  Our result reduces the
error bar by a factor 3 and removes a major source of the theoretical
uncertainty in $|V_{cb}|$.  This error can perhaps be further
decreased by choosing a different renormalization scheme, e.g. the
$V$-scheme.  This possibility will be examined in a future work.
However, with our estimate of the perturbative two-loop corrections to
$\eta_A$, the uncertainty in the zero recoil form factor is dominated
by the error in the $1/m_Q^2$ corrections; we adopt here the value
$\delta_{1/m^2}=-(5.5\pm 2.5)\%$ \cite{Neubert94,Neubert95} (for a
recent discussion of these corrections see \cite{Kapustin96}.)  Putting
this result together with our $\eta_A=0.960\pm 0.007$ we find for the
zero recoil form factor
\begin{eqnarray}
{\cal F}(1) =  \eta_A \left(1+\delta_{1/m^2}\right) = 0.907 \pm 0.026
\end{eqnarray}

We use the latest experimental data from the recent DELPHI analysis
for the $B^0\to D^{*-}l^+\nu$ decay rate.
We include the electroweak correction and the Coulomb enhancement, as
discussed after eq.~\ref{eq:spectrum}. They enhance the rate by
$1.3\%$ and $2.3\%$, respectively.  Altogether, we find 
\begin{eqnarray}
|V_{cb}| &=&
 0.0379 \pm 0.0020({\rm stat})
\nonumber \\
&& \pm 0.0025({\rm syst})
\pm 0.0011({\rm theory}).
\end{eqnarray}
We see that with the decreased theoretical uncertainty further
improvement in statistical and systematic accuracy can significantly
increase the precision of $|V_{cb}|$ and bring us closer to
overconstraining the unitarity triangle.

\acknowledgments 
I am grateful to Professor William Marciano for suggesting the
importance of electroweak and Coulomb corrections, and to Dr.~Dan
Pirjol for helpful discussions. I thank Professor J.H.~K\"uhn for his
interest in this work and support.  This research was supported by a
grant BMFT 056KA93P.


\begin{figure} 
\hspace*{0mm}
\begin{minipage}{16.cm}
\vspace*{5mm}
\[
\mbox{
\begin{tabular}{cc}
\psfig{figure=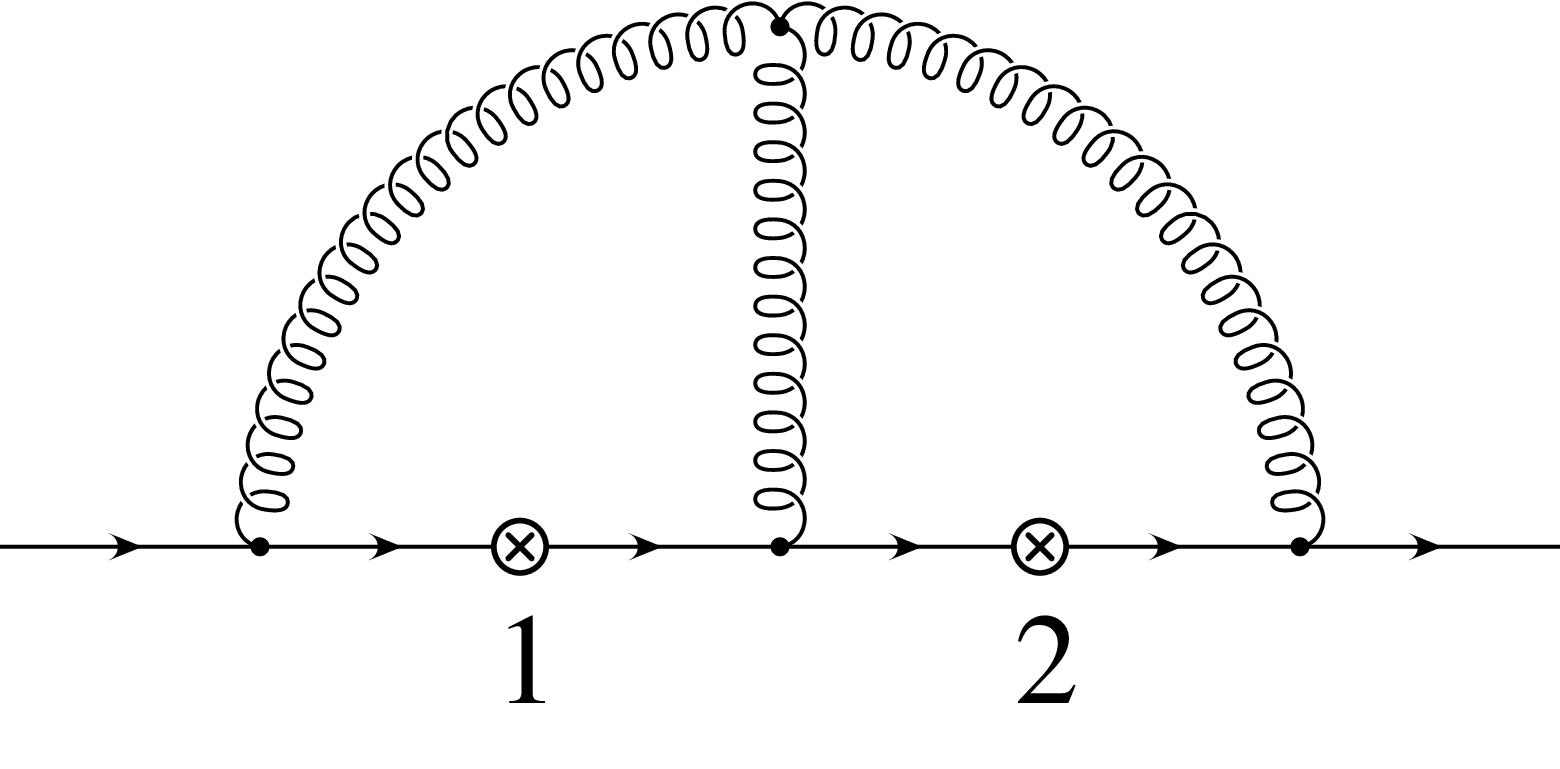,width=23mm,bbllx=210pt,bblly=410pt,%
bburx=630pt,bbury=550pt} 
&\hspace*{19mm}
\psfig{figure=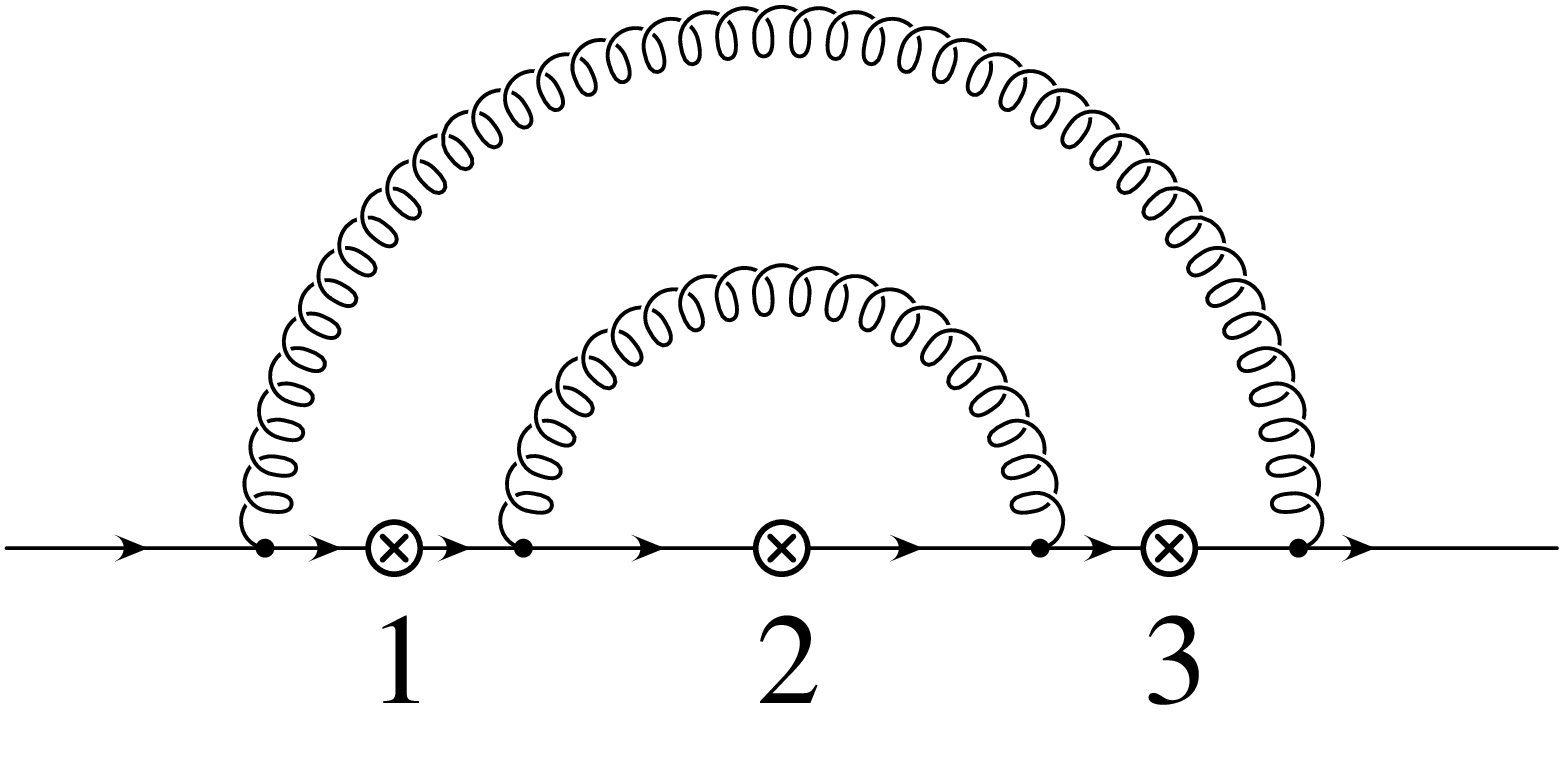,width=23mm,bbllx=210pt,bblly=410pt,%
bburx=630pt,bbury=550pt}
\\[5mm]
\hspace*{-13mm}(a) & \hspace*{7mm}(b) 
\\[10mm]
\psfig{figure=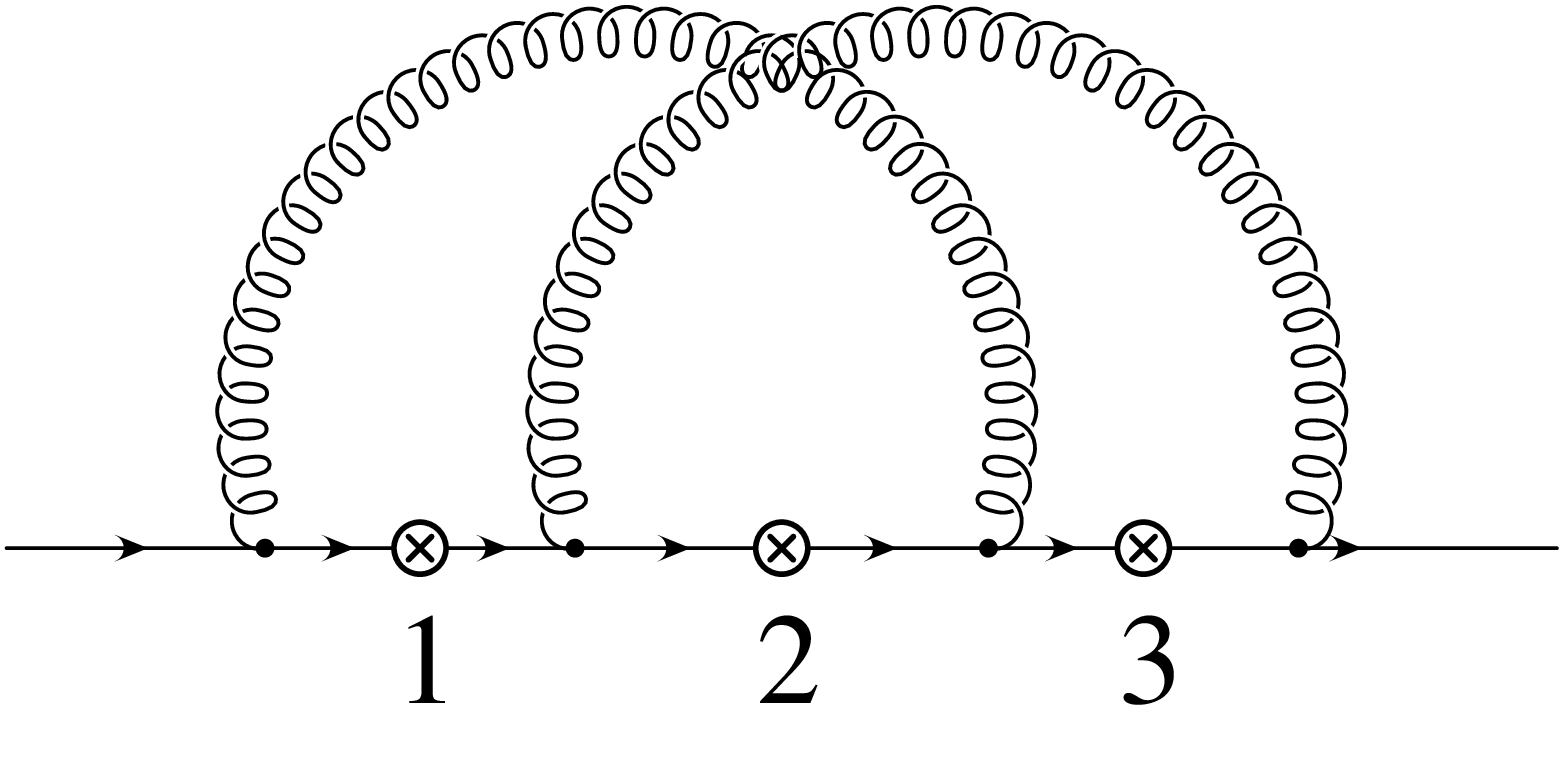,width=23mm,bbllx=210pt,bblly=410pt,%
bburx=630pt,bbury=550pt}
&\hspace*{19mm}
\psfig{figure=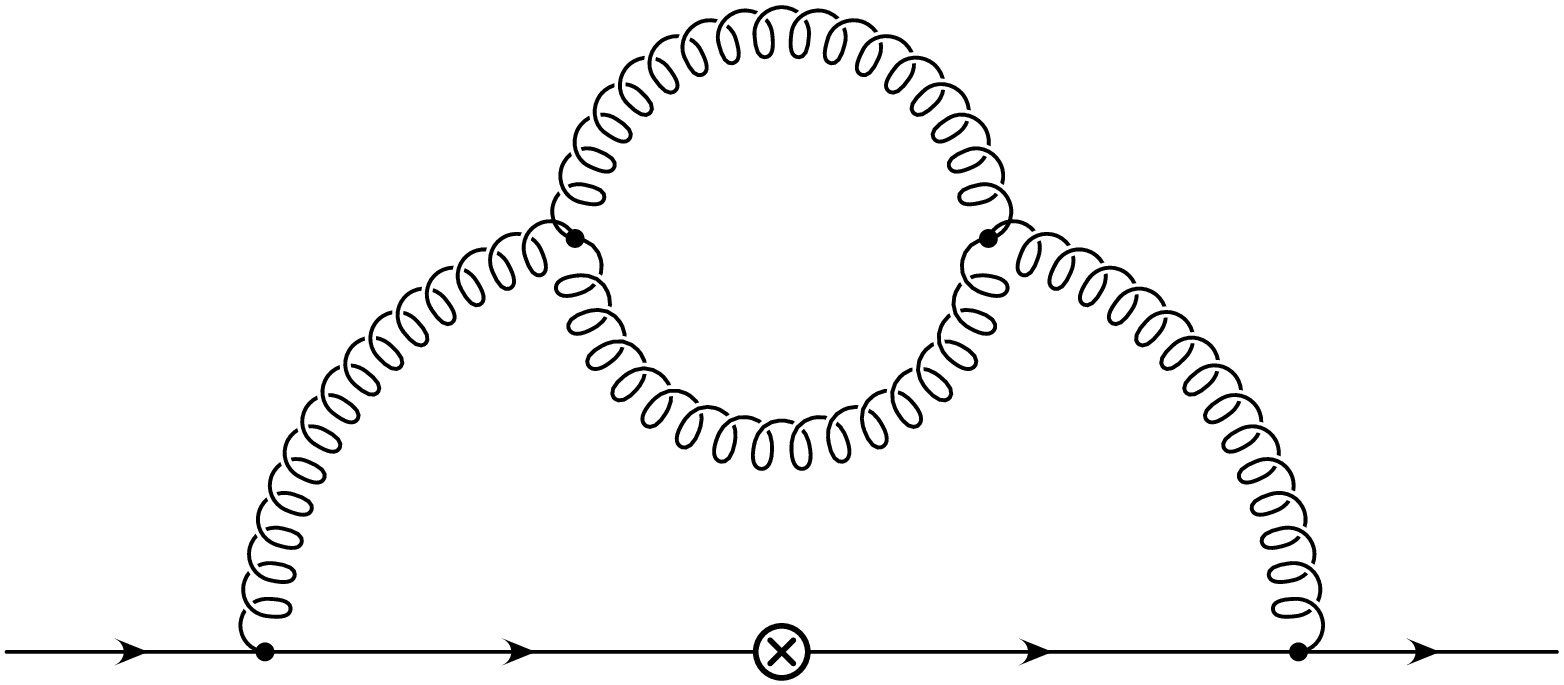,width=23mm,bbllx=210pt,bblly=410pt,%
bburx=630pt,bbury=550pt} 
\\[5mm]
\hspace*{-13mm}(c) & \hspace*{7mm}(d) 
\\[10mm]
\psfig{figure=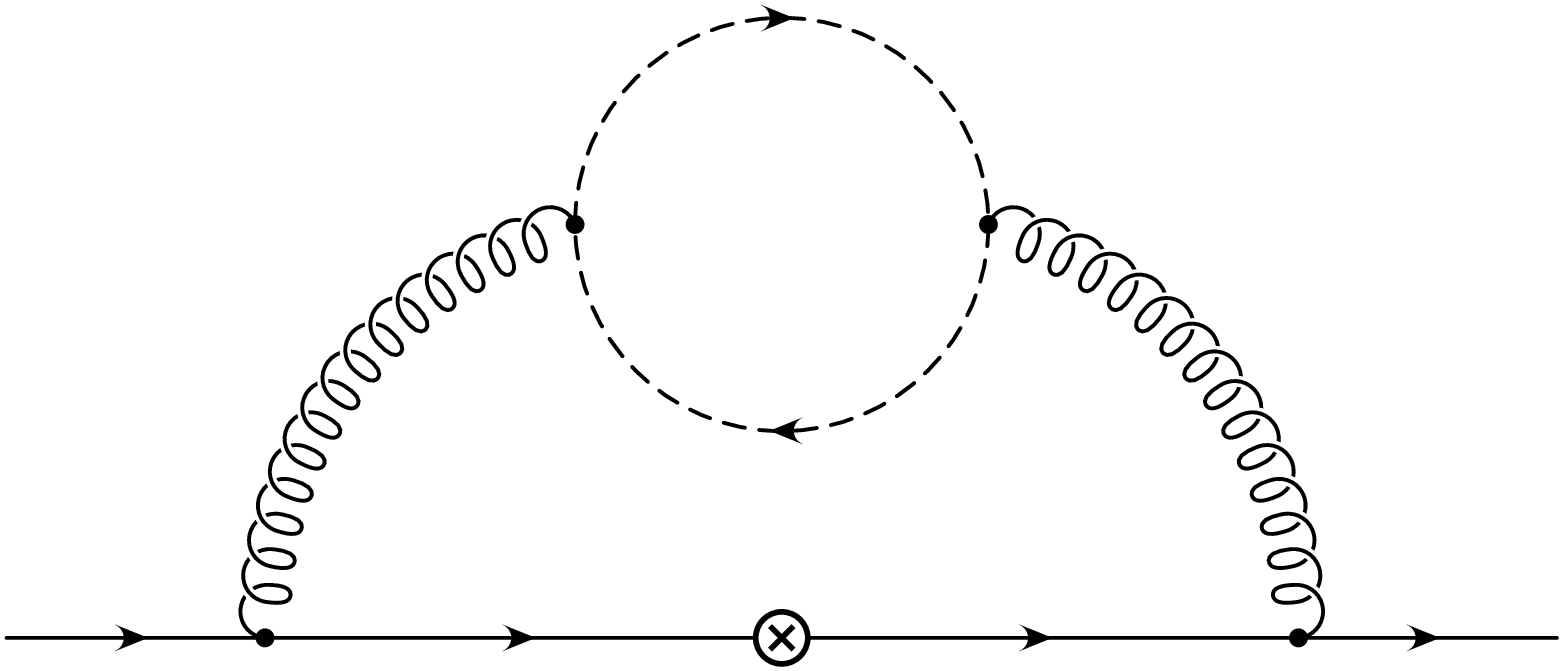,width=23mm,bbllx=210pt,bblly=410pt,%
bburx=630pt,bbury=550pt}
&\hspace*{19mm}
\psfig{figure=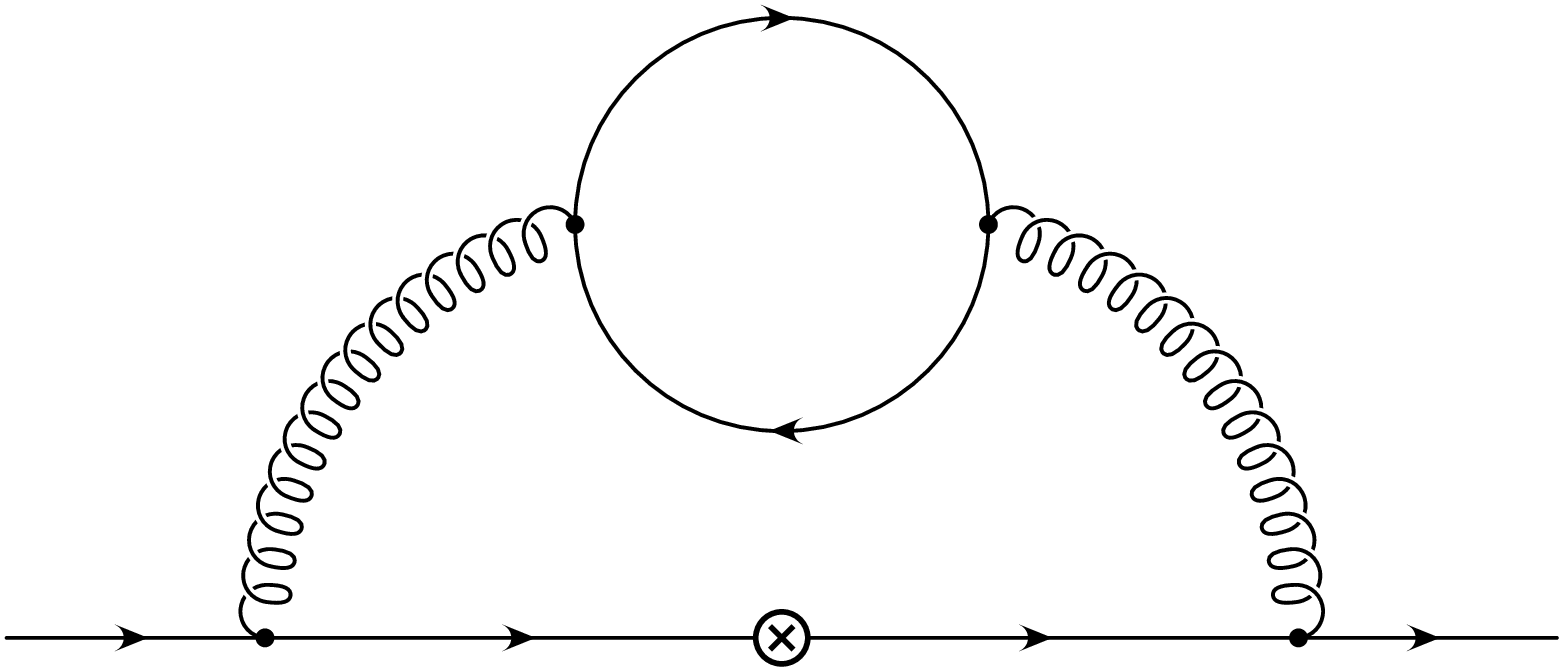,width=23mm,bbllx=210pt,bblly=410pt,%
bburx=630pt,bbury=550pt}
\\[4mm]
\hspace*{-13mm}(e) & \hspace*{7mm}(f) 
\end{tabular}}
\]
\end{minipage}
\caption{Two-loop QCD corrections to the $b\to c$ transitions at zero
recoil. Symbols $\otimes$ mark the coupling of the virtual $W$ boson
to the quark line. } 
\label{fig:twoloop}
\end{figure}

\end{document}